\documentclass[conference]{IEEEtran}
\IEEEoverridecommandlockouts
\usepackage{cite}
\usepackage{amsmath,amssymb,amsfonts, gensymb}
\usepackage{algorithmic}
\usepackage{graphicx}
\usepackage{textcomp}
\usepackage{xcolor}
\usepackage{listings}
\usepackage{courier} 
\usepackage[T1]{fontenc}
\usepackage{lmodern}
\usepackage{newtxtext,newtxmath}   
\pdfmapfile{+pdftex.map}

\usepackage[table,xcdraw,dvipsnames]{xcolor} 
\usepackage{listings}
\usepackage{courier}

\def\BibTeX{{\rm B\kern-.05em{\sc i\kern-.025em b}\kern-.08em
    T\kern-.1667em\lower.7ex\hbox{E}\kern-.125emX}}

\usepackage[left=0.75in,right=0.75in,top=0.75in,bottom=1.1in]{geometry}

\begin{document}

\title{C/N$_0$ Analysis-Based GPS Spoofing Detection with Variable Antenna Orientations 
}

\author{\IEEEauthorblockN{Vienna Li$^{1}$, Justin Villa$^{2}$, Dan Diessner$^{2}$, Jayson Clifford$^{2}$,  Laxima Niure Kandel$^{2}$}
\IEEEauthorblockA{$^{1}$Department of Electrical and Computer Engineering, Cornell University, Ithaca, USA \\
Email: vll24@cornell.edu}
\IEEEauthorblockA{$^{2}$Department of Electrical Engineering and Computer Science \& Center for Aerospace Resilience, \\
Embry-Riddle Aeronautical University, Daytona Beach, USA \\
Emails: villaj@my.erau.edu, \{cliffoj, niurekal, diessned\}@erau.edu}}

\maketitle

\begin{abstract}
GPS spoofing poses a growing threat to aviation by falsifying satellite signals and misleading aircraft navigation systems. This paper demonstrates a proof-of-concept spoofing detection strategy based on analyzing satellite Carrier-to-Noise Density Ratio (C/N$_0$) variation during controlled static antenna orientations. Using a u-blox EVK-M8U receiver and a GPSG-1000 satellite simulator, C/N$_0$ data is collected under three antenna orientations flat, banked right, and banked left) in both real-sky (non-spoofed) and spoofed environments. Our findings reveal that under non-spoofed signals, C/N$_0$ values fluctuate naturally with orientation, reflecting true geometric dependencies. However, spoofed signals demonstrate a distinct pattern: the flat orientation, which directly faces the spoofing antenna, consistently yielded the highest C/N$_0$ values, while both banked orientations showed reduced C/N$_0$ due to misalignment with the spoofing source. These findings suggest that simple maneuvers such as brief banking to induce C/N$_0$ variations can provide early cues of GPS spoofing for general aviation and UAV systems.
\end{abstract}


\section{Introduction}
Global navigation satellite systems (GNSS), particularly GPS, are critical to aviation for accurate positioning, navigation, and timing. However, GPS signals are inherently weak and follow predictable structures, making them vulnerable to spoofing attacks. GPS signals, which are composed of standardized Pseudo-Random Noise (PRN) codes and fixed ephemeris data, are susceptible to spoofing attacks. Common spoofing techniques include meaconing, which involves rebroadcasting real-sky (non-spoofed) signals with a delay, causing the receiver to compute incorrect positions without detecting a signal anomaly. More advanced spoofing involves using a software-defined radio (SDR) to transmit fabricated signals that are aligned in timing and Doppler shift to match the receiver’s expectation and gradually overpower legitimate signals.

In aviation, the implications of such spoofing are severe. The Flight Management System (FMS) typically prioritizes GPS as its primary navigation input, falling back on systems like the Inertial Reference System (IRS), DME/DME, and VOR/DME when GPS is unavailable. However, in recent incidents, spoofed signals have not only misled the FMS but also corrupted the IRS, which depends on GPS for periodic updates. When fed false position data, the IRS accumulates errors and loses integrity, while sensor fusion software misinterprets conflicting inputs from ground-based nav-aids and inertial sensors. This integrated navigation aid approach can result in the failure of navigation systems in some aircraft. 

Despite the growing threat, existing spoofing detection methods are often costly and rely on specialized hardware such as multi-element antennas or cryptographic signal real-sky authentication, limiting their applicability in general aviation \cite{bhatti2012gps}. The goal of this paper is to demonstrate a low-cost and effective method to detect spoofing by leveraging C/N$_0$ variations during controlled static antenna orientation. In this study, we examine how C/N$_0$ patterns change when the antenna is rotated among three fixed orientations. Under real-sky (non-spoofed) conditions, natural satellite diversity produces asymmetric C/N$_0$ changes with orientation. In contrast, spoofed signals---originating from a single fixed transmitter---result in the flat orientation receiving the highest C/N$_0$ values, with left and right banked positions showing reduced C/N$_0$s. This unnatural C/N$_0$ symmetry is used as a signature for identifying spoofing events.

The main contributions of this paper are summarized below: 
\begin{itemize}
    \item This paper presents a lightweight receiver-side GNSS spoofing detection technique that uses only standard C/N$_0$ measurements and satellite metadata, without requiring specialized antennas or hardware modifications.
    \item By leveraging the predictable geometric behavior of GNSS signals and identifying violations of this behavior, our approach enables spoofing detection that is:\\
a) \textbf{Robust} to signal content manipulation,\\
b) \textbf{Independent} of navigation solution anomalies, and\\
c) \textbf{Low-cost} and easily integrable into commercial GNSS receivers.
    
\end{itemize}
    

\section{Literature Review}
GPS spoofing detection techniques span signal analysis, sensor fusion, and machine learning. Angle-of-arrival (AoA) methods using off-the-shelf chipsets \cite{274743}, clock bias anomalies \cite{s23052735}, and RSSI-inferred movement \cite{9042947} have been effective in detecting inconsistencies from spoofed sources. Banking-induced signal variation in NGSO systems \cite{10328056} further supports spatial C/N$_0$ asymmetry as a detection metric. Broad surveys \cite{bhatti2012gps}, \cite{s20040954} emphasize the need for lightweight, hardware-independent solutions---especially for UAVs and civil GNSS receivers.

Sensor fusion methods using Kalman filters \cite{10223944}, AHRS accelerometers \cite{s20040954}, and motion units \cite{9968443}, \cite{10131729} have demonstrated strong spoofing detection performance. More recently, machine learning has enabled models based on Hilbert envelopes \cite{WANG2024103959}, representation learning \cite{article}, adversarial training \cite{s24186156}, and deep generative frameworks \cite{10495074}. These methods culminate in capsule-based networks for UAV spoofing detection \cite{khoei2025capsulenet}. However, many of these advanced detection mechanisms face limitations such as high computational complexity, reliance on extensive training data, and challenges in real-time deployment. This study contributes a simple alternative: identifying spoofing via consistent C/N$_0$ patterns across antenna orientations, requiring no additional hardware or cryptographic overhead.

\section{Experimental Setup}
Satellite data is collected using a u-blox EVK-M8U GNSS receiver, configured to log NAV-SAT messages at 5-second intervals over a 60-second duration per antenna orientation. Precisely cut foam blocks are used to simulate different antenna orientations and model the banking motion in an aircraft. A horizontally cut foam block is used to simulate a flat antenna orientation, and another block cut at a 45° angle is used to simulate left- and right-banked positions. This setup allows for consistent and repeatable positioning of the receiver.

To perform the spoofing experiments, a GPSG-1000 RF signal generator is used to simulate a spoofed GNSS constellation. The spoofed signal is generated using the daily almanac to maintain consistency with real-world satellite configurations, enabling controlled comparisons (Figure 1). The GPSG-1000 simulator produces a spoofed RF signal that is directly fed into the GNSS receiver’s RF input port through a calibrated coaxial cable. To guarantee complete isolation from real-sky (non-spoofed) satellite signals, all spoofing experiments are performed within a custom-built RF shielded enclosure lined with conductive mesh, effectively forming a Faraday cage. 

All data collected outdoors occurred under clear sky conditions to ensure optimal satellite visibility and signal quality (Figure 2). The Carrier-to-Noise Density Ratio (C/N$_0$) data for each satellite is extracted from the logs, timestamped, and grouped by PRN (satellite ID). Subsequent statistical analysis and graphical visualizations are performed using Python and Matplotlib. To analyze the spoofing effects, the signal strength profiles of real-sky (non-spoofed) and spoofed signals are compared across the different antenna orientations, and the patterns are used to identify inconsistencies that could be exploited to build the spoofing detector.
\begin{figure}[h!]
  \centering
  \includegraphics[width=0.4\textwidth]{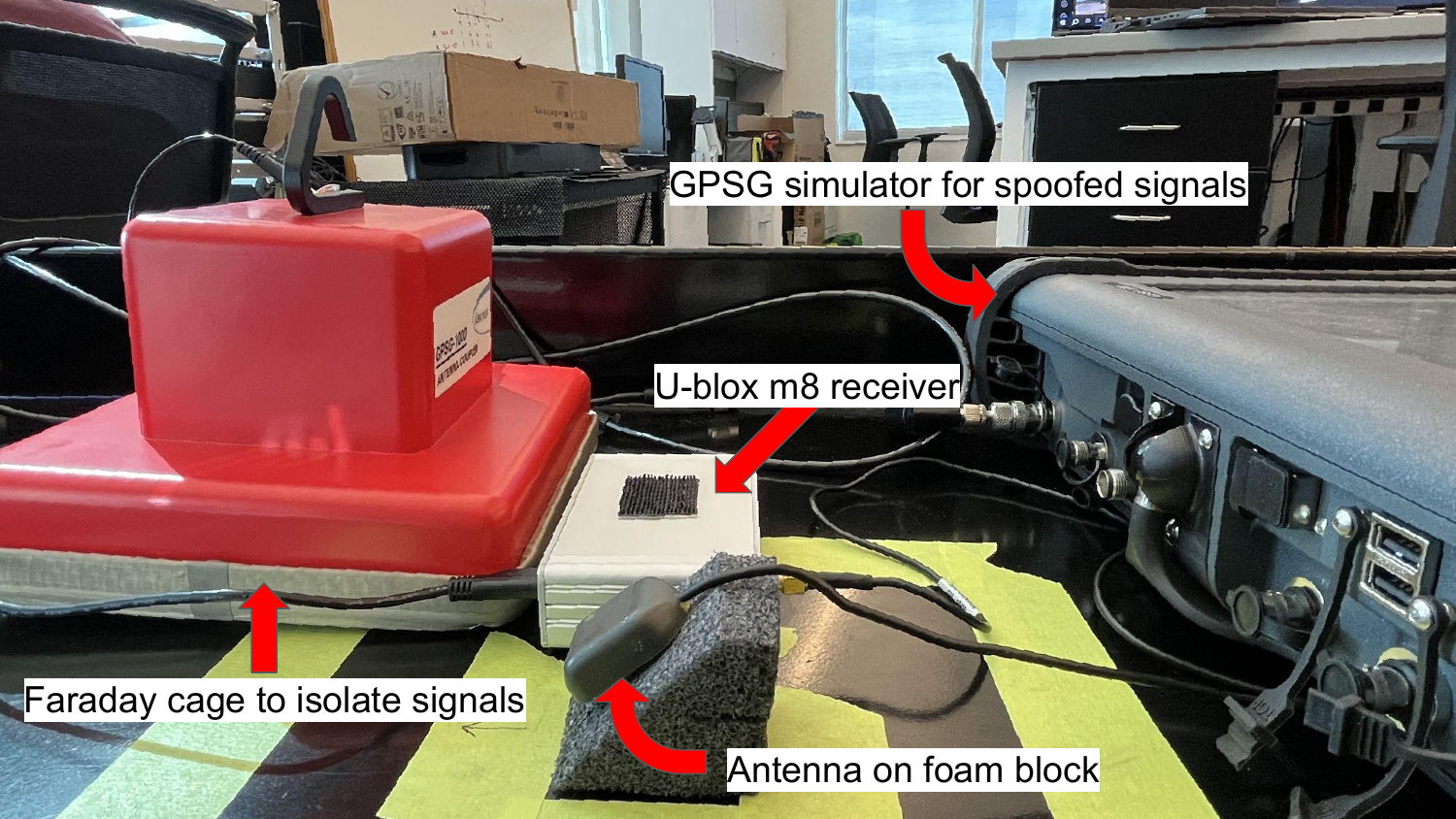}
  \caption{Spoofing detection experiment setup featuring the patch antenna, red Faraday cage for signal isolation, and GPS simulator on the right.}
  \label{fig:polar}
\end{figure}

\begin{figure}[h!]
  \centering
  \includegraphics[width=0.4\textwidth]{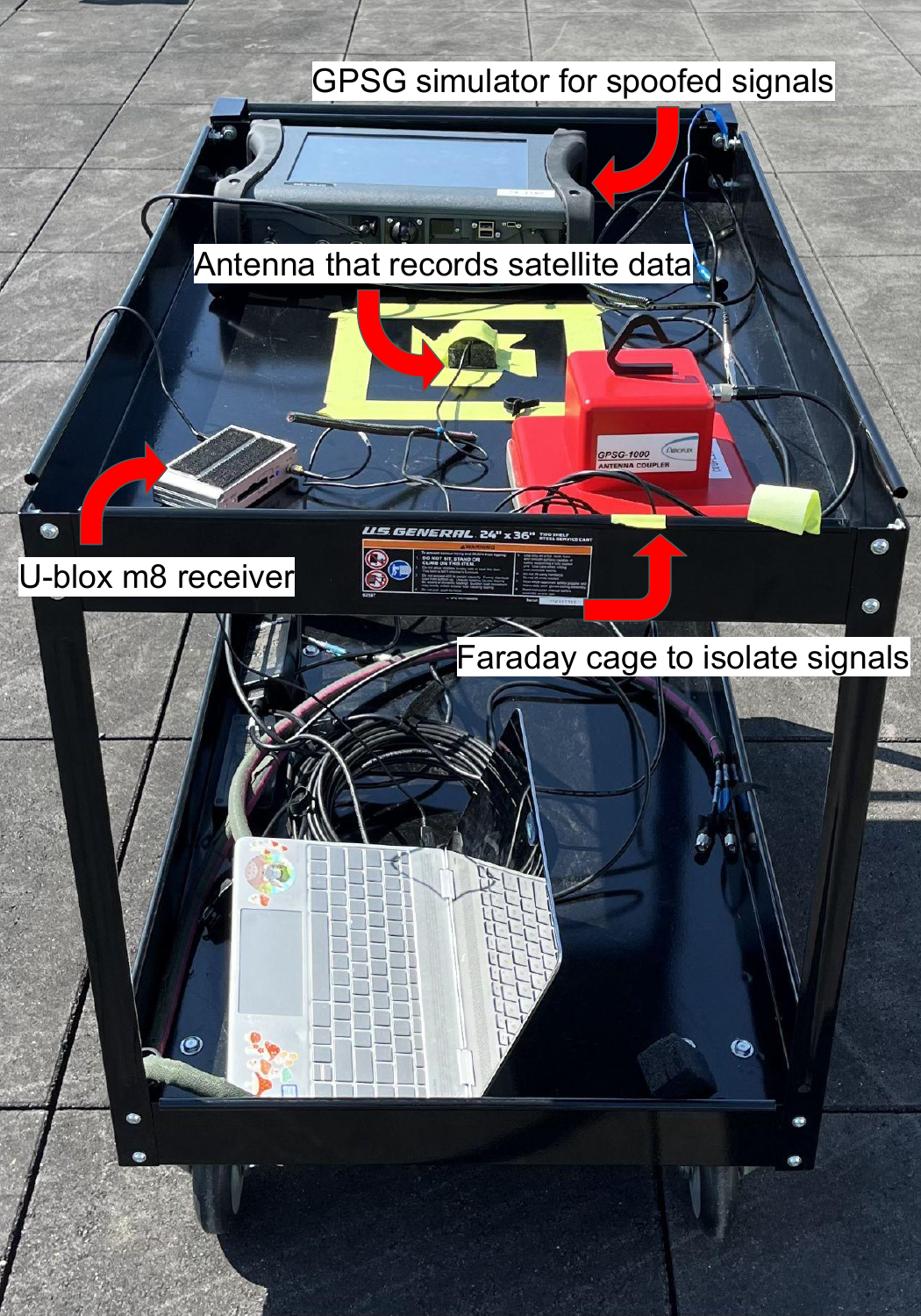}
  \caption{Outdoor data collection setup for capturing real-sky (non-spoofed) and spoofed GPS signals under controlled antenna orientations.}
  \label{fig:polar}
\end{figure}
\section{Methodology}
Building on previous studies done at the Embry-Riddle Aeronautical University Center for Aerospace Resilient Systems, this section describes the proposed methodology. In this study, we assume the noise power spectral density remains approximately constant across measurements since noise floor fluctuations are negligible compared to spoofing-induced C/N$_0$ changes in a shielded lab setup and short-duration experiments. Under this assumption, C/N$_0$ is proportional to signal strength, and thus variations in C/N$_0$ can be interpreted as analogous to variations in signal strength. With this assumption in mind, the primary goal is to establish a simplistic spoofing detection framework, leveraging the geometry of GPS signals and the banking angle of aircraft vehicles, since the spatial signal characteristics depend on the orientation of the antenna. By simulating left, flat, and right antenna banks, representative of the roll dynamics of an aircraft, the experiment models how C/N$_0$ changes with antenna orientation. The underlying hypothesis is that genuine GNSS signals will exhibit monotonic increasing and decreasing behavior in signal strength as the antenna banks are rotated but spoofed signals, being artificially generated from a static point source, lack this geometric consistency and will not demonstrate the same behavior. 

To validate this hypothesis, data sets are collected for each antenna orientation under both real-sky (non-spoofed) and spoofed signal conditions. These data sets are then parsed, pre-processed, and analyzed to identify characteristic trends in the signal strength per PRN (Pseudo-Random Noise code identifier). The analysis enables comparison of how real versus spoofed signals evolve across antenna orientations. Two spoofing detection models were developed to exploit these differences. The first is a rule-based approach, which manually labels PRNs as increasing or decreasing in signal strength within the non-spoofed dataset. Deviations from these expected behaviors are flagged in spoofed datasets, indicating possible signal manipulation. The second is a geometry-based model, which incorporates satellite azimuth, elevation, and antenna heading in the flat orientation to predict the expected signal strength trends. These geometry-driven predictions establish a baseline against which observed data can be evaluated, enabling automated spoofing detection through comparison of measured and expected trends.

\subsection{Data Collection}\label{AA}
To capture the effect of antenna orientation on satellite signal behavior, six distinct data collection scenarios are constructed, representing all combinations of signal source (non-spoofed or spoofed, and the three antenna orientation left bank, flat, or right bank). For the real-sky (non-spoofed) tests, the receiver is physically rotated and statically mounted in each orientation using foam blocks, enabling consistent simulation of an aircraft banking left, flying level, or banking right. For the spoofed experiments, identical orientations are recreated inside a shielded RF environment, but instead of receiving live satellite signals, the receiver is injected with simulated GPS signals from a GPSG-1000 RF generator. This yields six datasets in total: three from real satellite observations (\texttt{nonspoof\_left}, \texttt{nonspoof\_flat}, \texttt{nonspoof\_right}) and three from spoofed signal environments (\texttt{spoof\_left}, \texttt{spoof\_flat}, \texttt{spoof\_right}). Each dataset captures GNSS satellite signal parameters at 5-second intervals over a continuous 60-second period for each of the testing configurations. On average, each epoch includes data from approximately 12 to 14 unique satellites, identified by their satellite ID (\texttt{svId}).

The raw data is stored in comma-separated value (CSV) format with the following fields:

\begin{itemize}
    \item \textbf{\texttt{timestamp}}: Wall-clock time of the GNSS message (logged at 5-second intervals).
    \item \textbf{\texttt{svId}}: Satellite Vehicle ID (PRN number), which uniquely identifies each GNSS satellite.
    \item \textbf{\texttt{elev}}: Elevation angle in degrees above the horizon; higher values indicate satellites closer to zenith.
    \item \textbf{\texttt{azim}}: Azimuth angle in degrees from true north, defining the satellite's horizontal bearing.
    \item \textbf{\texttt{cno}}: Carrier-to-noise ratio in dB-Hz, a key indicator of signal strength and quality.
    \item \textbf{\texttt{qualityInd}}: A proprietary quality index provided by the u-blox receiver, used to indicate signal reliability.
    \item \textbf{\texttt{svUsed}}: A binary flag (1 or 0) indicating whether the satellite is used in the navigation solution.
\end{itemize}

The fields \texttt{svId}, \texttt{azim}, \texttt{elev}, and \texttt{cno} together describe which satellite is observed, its position in the sky, the strength of its signal, and whether it contributed to positioning. To ensure consistency across orientation tests, data is collected within carefully synchronized windows to maintain similar satellite visibility across all experiments, and the satellites' position is assumed to be static relative to the antenna.

\subsection{Data Visualization and Analysis}
After data collection, all six datasets are parsed and processed using Python. For each dataset, individual satellite records are grouped by their satellite vehicle ID (\texttt{svId}), effectively aggregating data per PRN across time. Since each dataset consists of repeated measurements over a 60-second period at 5-second intervals, multiple samples are available for each satellite in each orientation. The C/N$_0$ is a critical measure of the strength of the received signal and is averaged over time for each PRN within each dataset, resulting in an average C/N$_0$ value per PRN and orientation. These per-orientation averages formed the basis for trend comparison across the three antenna positions: left, flat, and right.

To visualize satellite positions, Matplotlib is used to generate polar plots based on the azimuth and elevation values associated with each PRN (Figure 3). These polar plots allow each satellite's spatial location to be mapped in a receiver-centric coordinate system, where the azimuth (angle from true north) defines the angular position around the plot and the elevation (angle above the horizon) determines radial distance from the center. Signal strength values (C/N$_0$) are then superimposed onto these plots using color coding and marker size, allowing for intuitive visual correlation between signal strength and satellite position.

\begin{figure}[h!]
  \centering
  \includegraphics[width=0.5\textwidth]{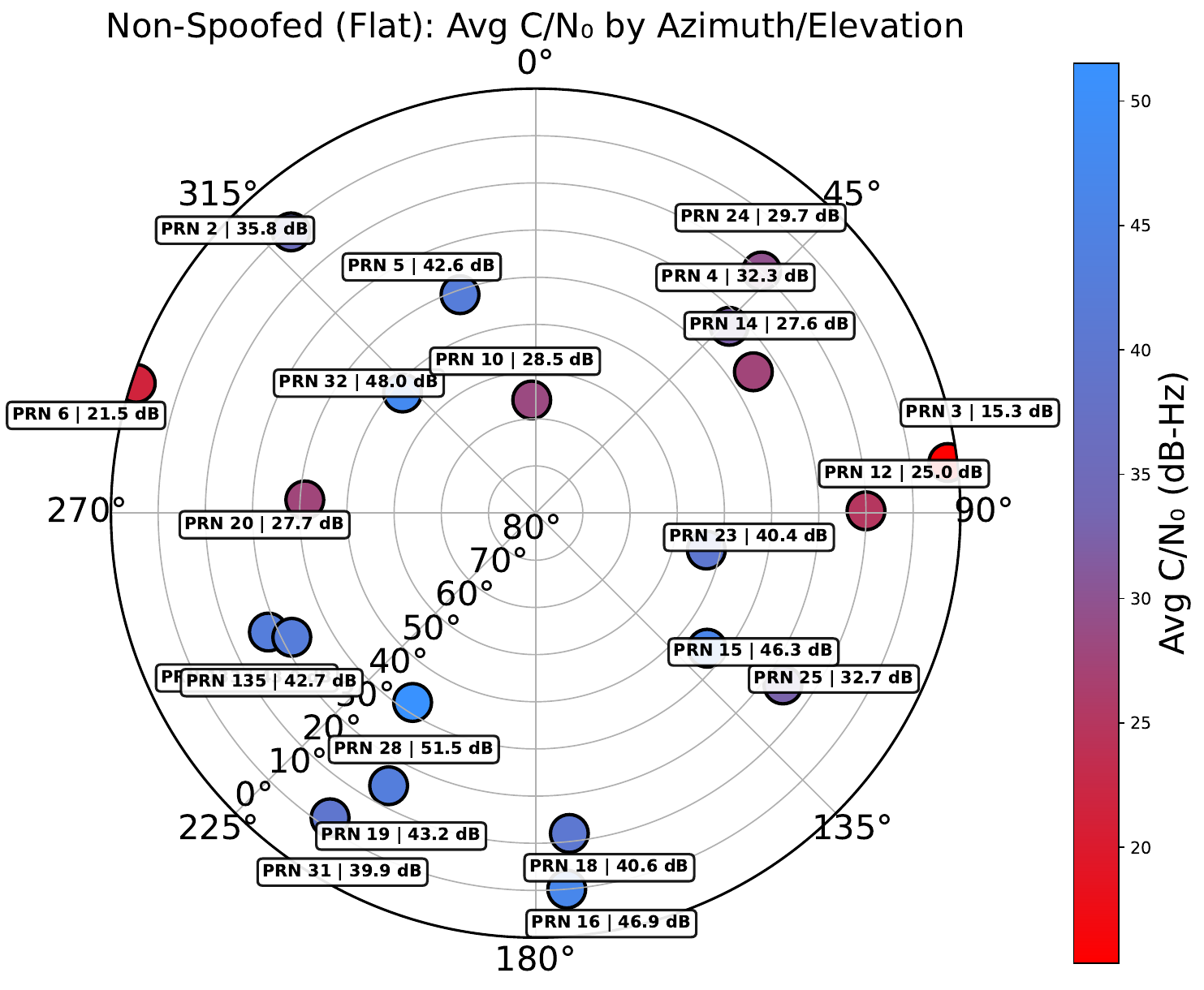}
  \caption{Non-Spoofed (top) vs. Spoofed (bottom) Signal Strengths for three orientations (left-banked, flat, and right-banked).}
  \label{fig:polar}
\end{figure}

Figure 4 reveals a clear geometric relationship in the real-sky (non-spoofed) datasets. As the antenna is physically rotated from the left $\rightarrow$ flat $\rightarrow$ to the right, the signal strength of each satellite changes in a predictable and spatially consistent manner. Satellites located in the left half of the sky plot (Figure 3) exhibit increasing signal strength trends across this rotation. In contrast, the satellites in the right half of the sky plot exhibit decreasing trends. 

For instance, satellites 5, 20, 32, 133, 135, and 138 (Figure 3) demonstrate a monotonic increase in average C/N$_0$ values as the antenna moved from left to right (Figure 4). This suggests these satellites entered the main lobe of the antenna's reception pattern, receiving a stronger signal coupling as the antenna rotated into alignment. Conversely, satellites 4, 15, 16, 24, and 25 (Figure 3), which are positioned in the hemisphere opposite to the antenna's facing direction, exhibited a clear decrease in signal strength across the same antenna motion, likely moving further outside the antenna’s optimal reception cone (Figure 5).

\begin{figure}[h!]
  \centering
  \includegraphics[width=0.4\textwidth]{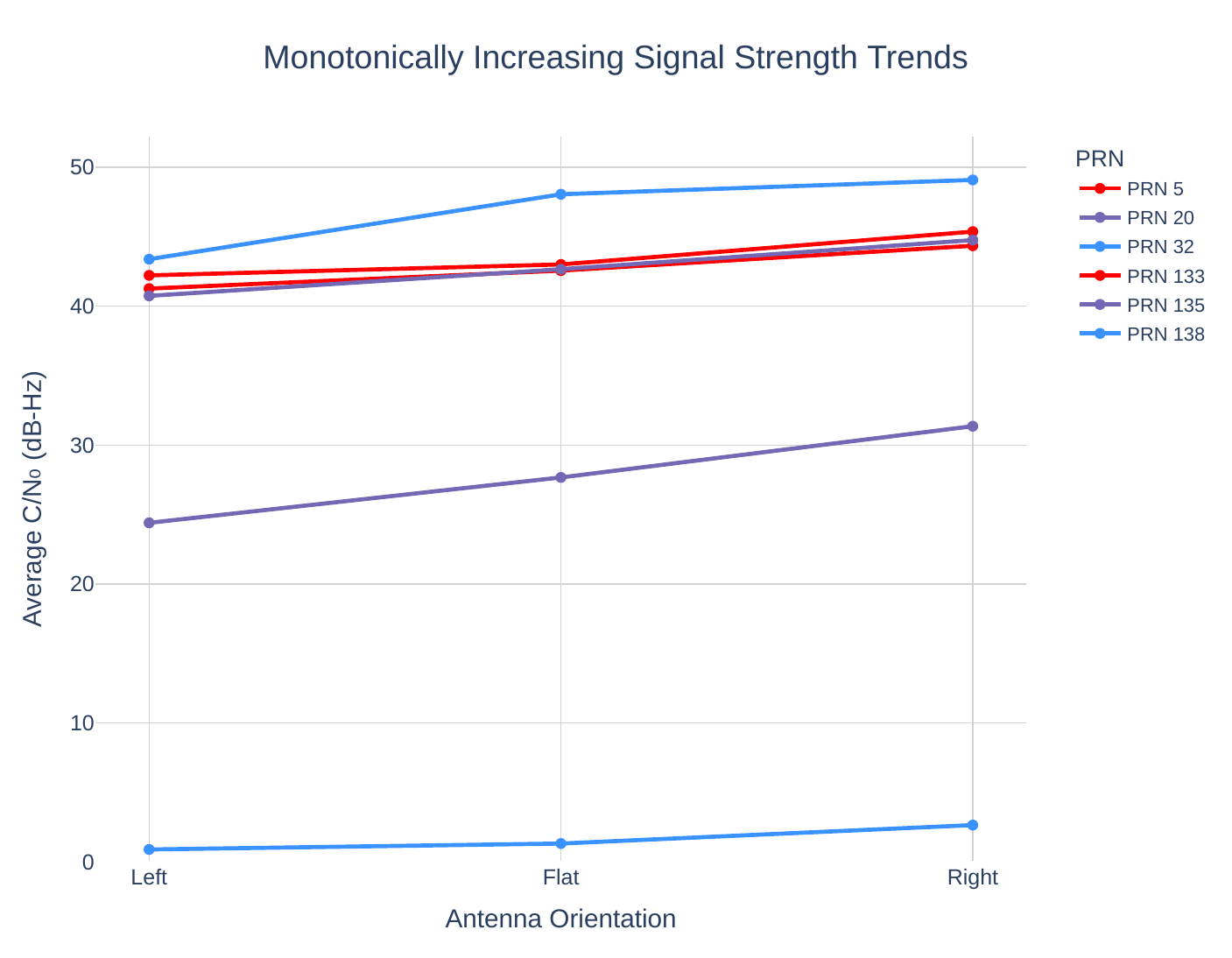}
  \caption{Plot showing satellites with increasing signal strength.}
  \label{fig:polar}
\end{figure}

\begin{figure}[h!]
  \centering
  \includegraphics[width=0.4\textwidth]{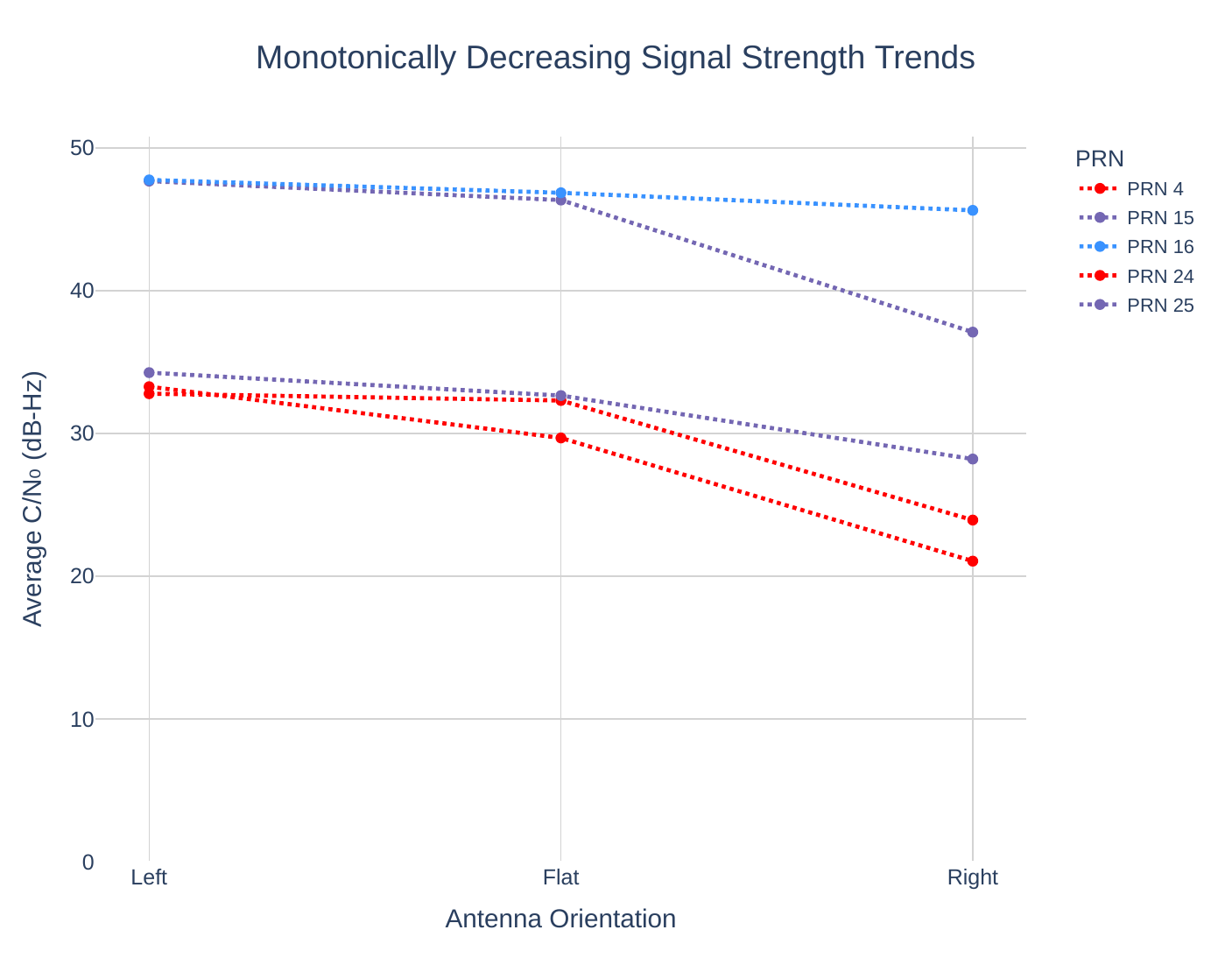}
  \caption{Plot showing satellites with decreasing signal strength.}
  \label{fig:polar}
\end{figure}

These signal changes are consistent and reflect the sensitivity of the antenna relative to the satellite constellation. The monotonic nature of these trends is a key characteristic of real-sky (non-spoofed) GNSS signals under a static, real-sky (non-spoofed) environment.

In contrast, spoofed datasets showed a breakdown of these trends (Figure 6). Since the spoofing signal originated from a single RF source and is injected directly into the receiver within a Faraday enclosure, it lacked any true spatial diversity. The signal strength values for spoofed PRNs showed a drastic difference compared to the normal data sets. Specifically, the signal strength is greatest at the center and falls for any bank direction because all spoofed satellite signals effectively emanate from the same point, rendering the antenna orientation irrelevant.  This lack of directional behavior, when compared with the well-structured trends of real GNSS signals, creates a strong contrast that could be exploited algorithmically for spoof detection.

\begin{figure}[h!]
  \centering
  \includegraphics[width=0.5\textwidth]{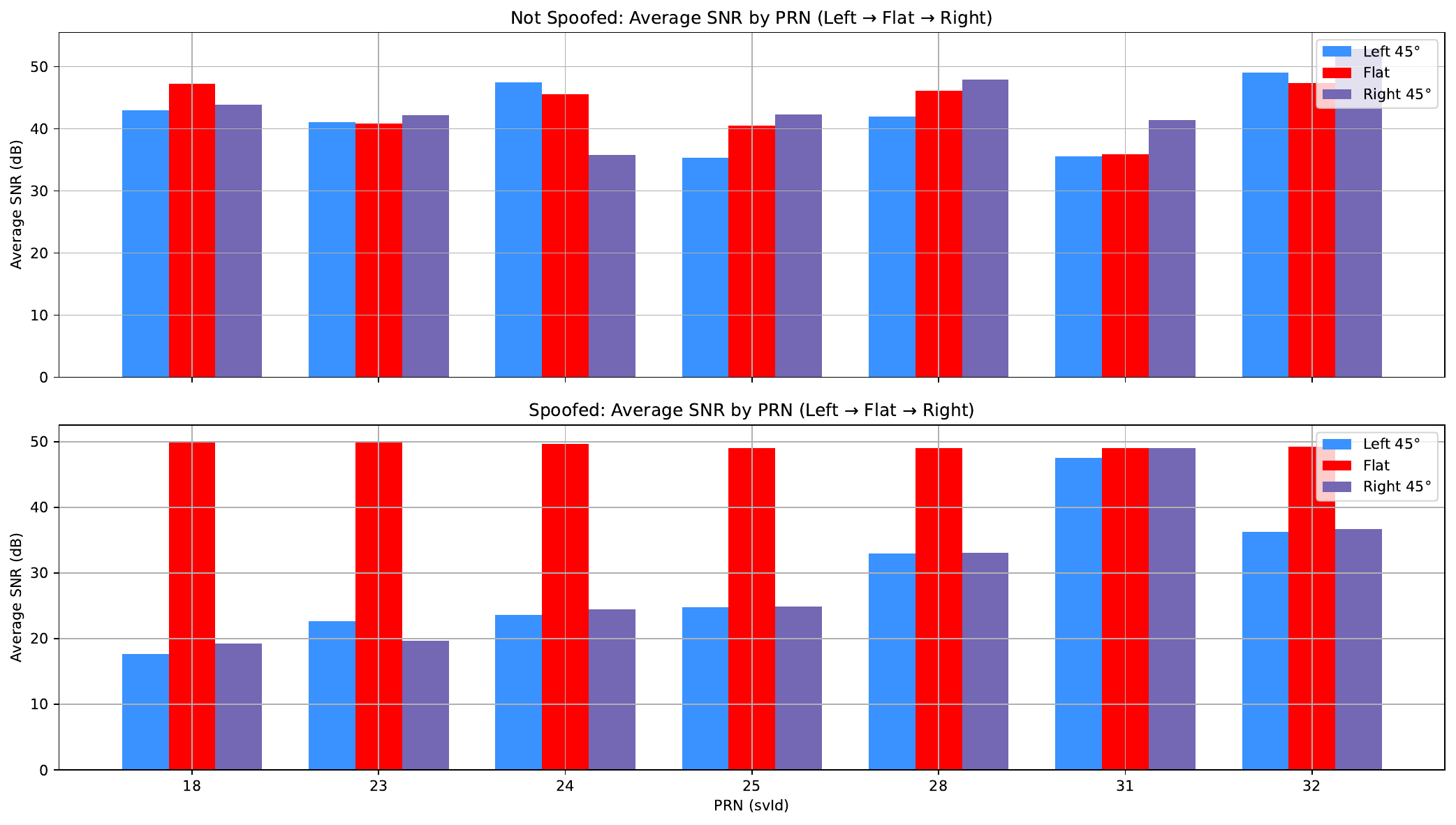}
  \caption{Carrier-to-noise ratio (C/N$_0$) measurements for spoofed and unspoofed signals across three antenna orientations: left-banked, flat, and right-banked.}
  \label{fig:trends}
\end{figure}

To quantify this contrast, signal trends are categorized by comparing the per-PRN average C/N$_0$ values across the three orientations. PRNs exhibiting strictly increasing or decreasing patterns are flagged as consistent and thus likely real-sky (non-spoofed), while PRNs showing flat or irregular trends are flagged as anomalous and potential indicators of spoofing. In addition to trend classification, variance metrics are computed to capture the degree of fluctuation in C/N$_0$ across orientations, further strengthening the ability to distinguish authentic versus spoofed signals.

Overall, the analysis of C/N$_0$ variation with respect to antenna orientation confirms the core hypothesis: real GNSS signals exhibit spatially coherent and orientation-dependent behavior, while spoofed signals lack this spatial diversity and fail to reproduce such patterns. These findings provide measurable features such as trend consistency and variance that serve as the foundation for the spoofing detection models developed in the next section.

\subsection{Building the Spoofing Detection Model}
Two spoofing detection approaches are implemented:

\begin{enumerate}
    \item \textbf{Rule-Based Detection Model} \\
    This model used the data from the non-spoofed (flat orientation) dataset and hard-coded expected trends into the model:
    \begin{itemize}
        \item Satellite azimuth and elevation values depicting expected monotonically increasing and decreasing signal strength are extracted from the non-spoofed baseline and then used to detect anomalies in the spoofed dataset. 
        \item The GPSG-1000 simulator is calibrated with the almanac immediately before the experiment begins to ensure a consistent sky plot and realistic simulation.
    \end{itemize}
    
    \subsection{Implementation Steps}

    \begin{enumerate}
        \item \textbf{Step 1: Load Data} \\
        Six CSV datasets are loaded into Python using \texttt{pandas}. These include spoofed and non-spoofed data across three antenna orientations (left, flat, right):

        \begin{lstlisting}[language=Python]
datasets = {
    "ns_left": pd.read_csv("3nl.csv"),
    "ns_flat": pd.read_csv("3nf.csv"),
    "ns_right": pd.read_csv("3nr.csv"),
    "s_left": pd.read_csv("3sl.csv"),
    "s_flat": pd.read_csv("3sf.csv"),
    "s_right": pd.read_csv("3sr.csv"),
}
        \end{lstlisting}

        \item \textbf{Step 2: Compute Avg C/N$_0$} \\
        For each dataset, compute the average C/N$_0$ per satellite (identified by \texttt{svId}):

        \begin{lstlisting}[language=Python]
def avg_cno(df):
    return df.groupby("svId")["cno"]
    .mean().to_dict()

avg_cno_data = {name: avg_cno(df) for name, df in datasets.items()}
        \end{lstlisting}

        \item \textbf{Step 3: Define Expected Trends} \\
        Satellites expected to have increasing or decreasing trends:

        \begin{lstlisting}[language=Python]
inc_prns = [5, 20, 32, 133, 138]
dec_prns = [4, 15, 16, 24, 25]
        \end{lstlisting}

        \item \textbf{Step 4: Trend Checks} \\
        Functions to check if values are strictly increasing or decreasing:

        \begin{lstlisting}[language=Python]
def is_inc(a, b, c):
    return a < b < c

def is_dec(a, b, c):
    return a > b > c
        \end{lstlisting}

\item \textbf{Step 5: Detect Violations} \\
Flag satellites whose signal trend violates expected monotonicity:

\begin{lstlisting}[language=Python]
def detect_violations(cno_left, cno_flat, cno_right):
    viol = []

    for prn in inc_prns:
        if prn in cno_left and prn in cno_flat and prn in cno_right:
            a, b, c = cno_left[prn], cno_flat[prn], cno_right[prn]
            if not is_inc(a, b, c):
                viol.append((prn, "inc", (a, b, c)))

    for prn in dec_prns:
        if prn in cno_left and prn in cno_flat and prn in cno_right:
            a, b, c = cno_left[prn], cno_flat[prn], cno_right[prn]
            if not is_dec(a, b, c):
                viol.append((prn, "dec", (a, b, c)))
    return viol
\end{lstlisting}

\item \textbf{Step 6: Run Detection} \\
Apply detection on spoofed data:

\begin{lstlisting}[language=Python]
viol_spoof = detect_violations(
    avg_cno_data["s_left"],
    avg_cno_data["s_flat"],
    avg_cno_data["s_right"]
)
\end{lstlisting}

\item \textbf{Step 7: Classify} \\
Classify based on the presence of violations:

\begin{lstlisting}[language=Python]
def classify(violations):
    return "spoofed" if violations else "non-spoofed"

print("Classification:", classify(viol_spoof))
print("Violations:", viol_spoof)
\end{lstlisting}
\end{enumerate}

  \item \textbf{Pattern-Based Detection Model} \\

Unlike the hard-coded rule-based model which uses fixed lists of satellites expected to increase or decrease signal strength, the pattern-based detection model dynamically predicts trends based on real satellite data:

\begin{itemize}
    \item The model utilizes azimuth and elevation information from the non-spoofed flat antenna orientation dataset (i.e., the polar plot data shown in Figure~\ref{fig:polar}.).
    \item Given an approximate antenna azimuth heading, the model calculates the antenna's left, flat, and right vectors. For each satellite (identified by PRN), the model computes its spatial vector and projects it onto the antenna vectors.
    \item This prediction step is summarized as follows:
\begin{lstlisting}[language=Python, basicstyle=\ttfamily\footnotesize]
mean = df.groupby("svId")[["azim", "elev"]].mean()
ant_vec_left = sph2cart((az-45) % 360, 0)
ant_vec_flat = sph2cart(az, 0)
ant_vec_right = sph2cart((az+45) % 360, 0)

for prn, row in mean.iterrows():
    prn_vec = sph2cart(row["azim"], row["elev"])
    if dot(ant_vec_left, prn_vec) < dot(ant_vec_flat, prn_vec) < dot(ant_vec_right, prn_vec):
        expected_increasing
        .append(prn)
    elif dot(ant_vec_left, prn_vec) > dot(ant_vec_flat, prn_vec) > dot(ant_vec_right, prn_vec):
        expected_decreasing
        .append(prn)
\end{lstlisting}

    \item The auto-detected expected increasing and decreasing PRNs from the dataset are:
    \[
    \text{Increasing: } [19, 20, 28, 31, 32, 133, 135, 138]
    \]
    \[
    \text{Decreasing: } [4, 12, 14, 15, 23, 24, 25]
    \]

    \item These predicted PRNs closely match the actual monotonic signal strength trends observed in the non-spoofed datasets, as visualized in Figure 3, with PRN satellites 4, 15, 16, 24, and 25 showing actual decreasing trends in the non spoofed dataset and PRN satellite 5, 20, 32, 133, 135, and 138 showing actual increasing trends in the non spoofed dataset.  Minor deviations between predicted and observed trends are primarily due to measurement noise, satellite geometry variations, and the limitations of the equipment. 
    \item This prediction now serves as the baseline, after which steps 5–7 follow the same procedure as the previous model: detecting violations, running the detection process, and classifying the anomaly.

\end{itemize}
\end{enumerate}

\section{Results and Conclusion}
To evaluate the proposed spoofing detection method, GPS C/N$_0$ values are collected under real-sky (non-spoofed) and spoofed conditions across three receiver orientations: flat, 45° bank left, and 45° bank right. Under real-sky (non-spoofed) conditions, C/N$_0$ patterns exhibited spatial asymmetry; specific satellites showed higher or lower signal strength depending on the banking angle. In contrast, spoofed conditions produced nearly flat C/N$_0$ profiles across all orientations, lacking the directional variation expected from the actual geometry of the satellite. This behavior confirmed the core hypothesis: real satellite signals produce orientation-sensitive C/N$_0$s due to their distinct positions in the sky, whereas spoofed signals, originating from a single fixed emitter, do not.

The analysis presented here emphasizes clear and repeatable C/N$_0$  patterns between spoofed and non-spoofed cases, rather than reporting formal detection metrics such as accuracy, false positive rates, or detection latency. Visual inspection of Figure 6 and the logged results shows that, in the real-sky datasets, per-PRN max–min C/N$_0$  values are typically on the order of several dB (the “typical” variation in signal strength per satellite is approximately 4–6 dB, with some PRNs showing variations up to 10–15 dB), whereas spoofed experiments exhibit much smaller per-PRN variation (roughly 0.5–1.5 dB, rarely exceeding ~3 dB). In other words, the change in C/N$_0$ between orientations is typically low in spoofed tests, while real-sky scenarios show changes show drastically higher variation. This measurable discrepancy provides a reliable signature for the detection of spoofing as it effectively distinguishes spoofed environments without requiring cryptographic signals, IMUs, or machine learning models, making it lightweight and highly deployable.

\section{Limitations and Future Work}

While the proposed detection method demonstrates promising results, it also presents some limitations. The current setup demonstrated in this paper relies on a static test configuration with discrete antenna orientations, rather than conducting measurements with the antenna in continuous motion. However, the method is expected to be especially well suited for dynamic platforms such as small UAVs, where pitch and roll angles vary more significantly in continuous motion, and as such the trends should be more pronounced. Although larger aircraft have slower attitude changes, these challenges can be mitigated using dual-receiver or dual-antenna configurations already employed for redundancy and coverage. With two antennas at different fuselage locations, real satellite signals will produce distinct C/N$_0$ responses at the two receivers due to their different lines of sight. This contrast enables detection of spoofing even in steady, level flight, thereby extending the applicability of the method from agile UAVs to large commercial transport aircraft. 

Future work will focus on dual or multi-antenna configurations, combined with continuous real time detection. Validation in dynamic flight environments will also be pursued, leveraging onboard avionics data and inertial measurement units (IMUs) to track antenna orientation in real time. Additionally, testing under more complex spoofing scenarios (e.g., mobile or multi-source spoofers) and diverse environmental conditions (e.g., urban canyons) will be essential. Long-term goals include real-time implementation on embedded platforms and integration with multi-modal spoofing detection frameworks combining C/N$_0$, timing, and inertial data.

\section{Acknowledgment}
This material is based upon work supported by the National Science Foundation award CNS-2244515.

\end{document}